\documentclass[3p,times,twocolumn,procedia]{elsarticle}

\usepackage{ecrc}

\volume{00}

\firstpage{1}

\journalname{Nuclear Physics B Proceedings Supplement}

\runauth{Ya.I. Azimov}

\jid{nuphbp}

\jnltitlelogo{Nuclear Physics B Proceedings Supplement}

\usepackage{amssymb}

\usepackage[figuresright]{rotating}

\begin{document}

\begin{frontmatter}



\dochead{}

\title{Unexpected Mesons $X, Y, Z$, ...  \\(tetraquarks? hadron molecules? ...)}


\author{Ya.I.~Azimov}

\address{Petersburg Nuclear Physics Institute \\
Gatchina 188300, Russia}

\begin{abstract}
This talk briefly discusses the set of meson resonances
discovered in the latest decade. They are frequently treated
in the literature as tetraquarks or hadron molecules. Our
consideration (using the energy-time uncertainty relation)
suggests, however, that the most reasonable description for
each of them may be a two- (or more-) component Fock column,
with one line being a conventional quark-antiquark pair, and
the other line(s) corresponding to a charmed (or beauty)
meson-antimeson pair near its threshold. Detailed investigation
of decay properties might allow to reveal presence of several
Fock components and separate their~contributions.

\end{abstract}


\end{frontmatter}


\section{Introduction. Reminder}
One of the basic points of the Standard Model is the existence of quarks
(and, of course, corresponding antiquarks) of 6 different kinds (flavors)
with different masses: 3 quarks $u,c,t$ having the electric charge $+2/3$
and 3 quarks $d,s,b$ with the charge $-1/3$. Further, it is widely
accepted at present that every (at least well-studied) meson consists
of one quark and one antiquark having, generally, different flavors. There
is only one exception here: the heaviest $t$-quark is so short-living that
it has not enough time to produce any hadrons.

Such a simple picture provides limitations for quantum numbers of both
mesons and baryons. For instance, baryons can not have positive
strangeness $S$, or isospin $I$ higher than 3/2; mesons can not have $I$
higher than 1, their $S$ may be only $0,\pm1$. Limitations arise also
for $J^{PC}$ values: $q\bar{q}$ mesons can have, \textit{e.g.},
$J^{PC}=0^{-+}$, but can not have  $0^{+-}$.

Note, however, that such a picture is inconsistent with any version
of relativistic field theory, where one can not exclude presence of an
arbitrary number of virtual quark-antiquark pairs and/or gluons.
Therefore, adequate description of any hadron should use a Fock column,
where lines correspond to particular configurations (but with the same
``global'' quantum numbers, like $I,J, P, C,$ and so on).

When describing the lower hadron states by the constituent quark model,
the model parameters and interaction potentials are usually chosen so
that higher Fock components can be ignored, with some accuracy. It is
not evident, however, whether the same approach, with the same parameters
and potentials, may be applicable for excited states as well.

\section{``Prehistoric'' tetraquarks}

Up to now, there have been found no mesons with ``non-canonical'' quantum
numbers, which could not exist in a quark-antiquark system. Nevertheless,
there is a tendency to explain any anomalous features of a meson resonance
by presence, or even dominance, of tetraquark (\textit{i.e.,} two
quark-antiquark pairs) component(s) in its structure. As an alternate, one
could consider hadron molecules, bound states of two (or more) hadrons.
For the first time, such hypotheses were applied to the scalar mesons
$a_0(980)$ and $f_0(980)$ (the corresponding story is briefly presented
in the Introduction to Ref.\cite{sc-mes}). Though the discussions are still
continuing, the necessity of the tetraquark nature of the scalar mesons
stays unproven (see, \textit{e.g.}, Conclusions in Ref.\cite{sc-mes}).

Later, the tetraquark configurations were searched for in excited
$D$-mesons, such as $D(2400)$ and heavier. The corresponding discussions
still arise again and again, but any definite conclusion has not been
reached yet, just as for the scalar mesons.

\section{New era}

New, and numerous, class of states appropriate to search for tetraquark
configurations was revealed about ten years ago, initially in studies at
$B$-factories. They are denoted by the symbols $X, Y, Z$. These states are
mainly found in $B$-meson decays, but also as a part of cascades in the
process $e^+e^- \to \mathrm{hadrons}$. Their common property today is
presence of the charmed $c$-quark pair among decay products. The charm may
be hidden, as a charmonium state, or open, as a charmed meson pair (there
have been found also several states with the $b$-quark pair in decay products).

The whole list of observed new states and their decay modes may be found
in a number of reviews on this new spectroscopy (some of more recent talks
and papers see, \textit{e.g.}, in Ref.\cite{revs}). Even more papers and
talks discuss the $X, Y, Z$ studies at particular experimental facilities.
All observed states of this group are included also in the Report of Particle
Physics~\cite{RPP}. Note, however, that those states are usually seen as
peaks in the mass spectra of decay products. Meanwhile, as known, peaks
might arise not only due to resonance states, but also because of some
kinematic effects. It is important in this respect, that the resonance
character of the peak has been demonstrated for at least one of $Z$ states,
$Z(4430)$, by the energy dependence of the phase of the corresponding
amplitude~\cite{lhcb}. All other peaks $X,\, Y,\, Z\,$ are only supposed today
to be true resonances (see, however, more general theoretical arguments
against pure kinematic effects in Ref.\cite{poles}).

But even if all those states were proved to be just resonances it would not
mean that we understand their internal structure. Each of them might consist
of a familiar quark-antiquark pair, or be a more complicated system. The
most frequent pictures considered in the literature are tetra quarks, bound
states consisting of two quarks and two antiquarks, or hadron molecules,
bound states of two (or more) hadrons. Clear physical discrimination of
these pictures is usually not discussed, but methods of calculation may be
different. For instance, tetraquarks are frequently described by sum rules,
though various other approaches have been applied as well. Many publications
pretend to give satisfactory description of a particular state, but nobody
could present a picture of the $X, Y, Z$ states as a whole. Thus, their
theoretical status stays uncertain. Representative, in this respect, is the
fate of Ref.\cite{sasa}. Its first version was able to see one $Z$ state, as
a result of lattice calculations, while its second version sees no such states
in the whole investigated region.

\section{Mesons $X$ and $Y$}

All known states of these kinds are neutral. Being tetraquarks (with one pair
$c\bar{c}$) or hadron molecule (of two charmed mesons), they could have isospin
$I=0$ or 1. Decay modes do not give a definite answer. For instance, $X(3872)$ can
decay both to $\rho^0 J/\psi$ (with $I=1$ in the final state) and to $\omega
J/\psi$ (with $I=0$), which means, of course, isospin violation in one of the
channels. Were the $X(3872)$ isovector, it should have a charged companion.
However, despite intensive searches, no charged companions have been found for
any of $X, Y$ states. This favors $I=0$ (if so, isospin violating should be the
decay to $\rho^0 J/\psi$).

A meson with such quantum numbers and with charmed quark pair (open or hidden)
in all decay modes could be just an excited state of charmonium. Indeed, the
measured $X, Y$ masses are close enough to calculated (alas, model-dependent)
levels of charmonium. For one of such mesons, $X(3915)$, the Particle Data
Group have even made their minds to identify it with a charmonium level,
$\chi_{c0}(2\,{}^3P_0)$~\cite{RPP}. However, calculated (again, model-dependent)
decay properties differ from the observed ones (for the particular case of
$X(3915)$, see Ref.\cite{ols2}). Thus, identification of $X, Y$ mesons as
charmonium levels stays questionable, though admissible.

\section{Mesons $Z$ }

A harder problem is the nature of states $Z$. They have neither strangeness,
nor charm, nor beauty, but all their decay modes, observed up to now, contain
open or hidden charm (for $Z_c$ states; there have been found also a couple
of states $Z_b$, which decays produce open or hidden beauty; this makes the
problem more general). However, all the $Z$ states are charged and have
isospin $I=1$ (for one of them, $Z_c^{\pm}(3900)$, the neutral component,
$Z_c^0(3900)$, has been found as well~\cite{neutr}). Therefore, they
definitely can not be charmonium (or bottomonium) levels. On the other hand,
all the $Z$ states have rather large widths (typically, some tens of MeV or
even more). Thus, their decays are governed by strong interactions, which
conserve every flavor. Therefore, the final $c$- or $b$-quarks could not arise
as a result of flavor changes in the decay (as could be in weak decays). On
the other hand, pair production of heavy flavors should be suppressed,
according to the Zweig-Iizuka rule. These facts provide a hint of presence
of the heavy quark pair just in the initial state. Were the hint true, the
$Z$ states would indeed be tetraquarks (\textit{e.g.}, $u\bar{d}c\bar{c}$)
or hadron molecules with the same quark content. Such situation is believable,
but is not proven yet and stays uncertain. In any case, nobody could achieve
general description of the $Z$ states along such lines.

\section{Possible role of thresholds }

In the space-time picture, higher Fock components are related to quantum
fluctuations. When considering hadrons in QCD, those fluctuations may be
described either as containing additional $q\bar{q}$ pairs and/or gluons
(compare to virtual photons and/or $e^+e^-$ pairs in QED), or as two- or
multihadron systems. The fluctuations arise and then disappear after some
characteristic time. According to the well-known energy-time uncertainty
relation, the lifetime of a fluctuation (in terms of virtual hadrons) is
the shorter, the larger is difference between the initial state mass and
physical mass of the virtual hadronic state.

If a hadron under consideration is a resonance state, it has itself some
finite lifetime. Let us assume that the resonance has just a canonical
quark-antiquark pair as the basic Fock component. Now, if a fluctuation
develops a (virtual) hadron system with its physical mass far above the
resonance mass, then the fluctuation has very short lifetime. It arises
and disappears before the resonance decays. Of course, such fluctuation
affects properties of the resonance, but only as a correction. If, just
opposite, the resonance mass is far above the threshold of a hadron system
in fluctuation, then the arising hadrons turn out to be real. They rapidly
run away after the fluctuation has arisen. Sure, the resonance lifetime
in such a case is mainly related not to the time of running away, but to
the time (and probability) of producing the corresponding fluctuation.

These two extreme considerations show that thresholds may play a special
important role. Indeed, if the fluctuation produces a hadron system in a
limited mass range near its threshold, then the fluctuation lifetime may be
near or even longer than the resonance lifetime. In such a case, when
describing the resonance, one should consider the fluctuation as permanent.
In other words, a near-threshold fluctuation can not be averaged out, it
becomes effectively enhanced and ``stabilized''. The corresponding Fock column
for the resonance can not be considered as one-component; even minimally,
it should be two-component. Of course, presence of additional Fock components
should be accounted for when calculating masses and decay properties of the
resonances.

Phenomenologically, this situation arises if the resonance Breit-Wigner peak
overlaps a threshold. To some extent, the case is similar to the known cusp
effect, which provides enhancement of an elastic cross section in a narrow energy
range near an inelastic threshold (discussion of the $Z$-states in respect with
cusps may be found in Ref.\cite{cusp}). Just as for cusps, the largest contributions
to the higher Fock components should come from hadron pairs in the near-threshold
$S$-wave state.

Among the "old" resonances, some cases seem to encounter just such threshold
effects. They are, first of all, the scalar resonances $f_0(980)$ and $a_0(980)$,
mentioned above as "prehistoric" tetraquarks. Their masses and widths~\cite{RPP}
are such that the Breit-Wigner peaks overlap the $K\bar{K}$ threshold. Moreover,
the resonance masses are so close to the kaon-pair threshold(s) that the resonance
properties are affected by the difference of thresholds for the charged and neutral
kaon pairs. This distorts isotopic relation between kaons and produces apparent
isospin violation, thus generating the observable $f_0$-$a_0$ mixing.

The other interesting example is the hyperon resonance $\Lambda(1520)$. Its decay
to $\Sigma(1385)\pi$ has, formally, no energy release and, therefore, vanishing
final-state phase space. At first sight, it should be kinematically forbidden
or, at least, suppressed. However, the branching ratio for the mode
$\Lambda(1520)\to\Sigma(1385)\pi$ is unexpectedly large, about 10\%~\cite{RPP}.
It could be just a result of the threshold enhancement.

This discussion shows that higher Fock components can not be completely expelled.
One may be able to construct a model and adjust its parameters so to describe the
lowest hadron states (nearly) without higher components. But for excited states,
those components, most probably, will occur non-negligible. Of course, such
expectations are equally applicable to both mesons and baryons.

Note that the arising picture differs from the two approaches most popular in the
literature. For instance, a meson with such structure is not a canonical tetraquark,
since it contains a quark-antiquark component. Hence, it may have only quantum
numbers which are admissible for the quark-antiquark pair; in terms of the $SU(3)_F$,
it can belong only to the lowest multuplets. On the other side, the hadron component
of such a meson is not a bound molecule, it is dominated by the hadron near-threshold
system.

\section{New mesons as Fock columns }

Let us apply the above viewpoint to the $X(3872)$, with mass $3871.7$~MeV, width
$<\!1.2$~MeV, and $J^{PC}=1^{++}$~\cite{RPP}. We assume that its lowest component
corresponds to $c\bar{c}$ with $I=0$. Then, it overlaps the threshold of
$D^{\,0}\bar{D}^{*\,0}$ (or $D^{*\,0}\bar{D}^{\,0}$) at 3871.8~MeV, but not the
threshold of $D^+\bar{D}^{*-}$ (or $D^-\bar{D}^{*+}$) at 3879.9~MeV~\cite{RPP}.
Note that the $X(3872)$, because of its spin-parity, is not coupled to $D\bar{D}$.
Thus, the corresponding Fock component can contain only neutral $D$-meson
contribution (in terms of the quark content, it may be considered as the
tetraquark configuration $\bar{u}u\bar{c}c$). Evidently, it is composed of two
isospins, $I=0$ and $I=1$, with equal intensity. This may explain why the decays
of $X(3872)$ to $J/\psi \rho$ and $J/\psi \omega$ have near the same branchings,
if these decays go mainly through rescattering of $D$-mesons. On the other side,
radiative decays to charmonium states could be mainly determined by the
$\bar{c}c$ component. Thus, accurate studies of the resonance decays may reveal
presence of several essential Fock components and separate their contributions.

As another example, let us consider a particular $Z$-state, $Z^\pm(4430)$ with
$J^P=1^+$, observed in decays to $\psi^\prime\pi^\pm$ (experimental information
on this state is briefly reviewed in Ref.\cite{lhcb}). It has mass about 4480~MeV
and width about 200~MeV~\cite{lhcb, belle}. Evidently, its main decays should be
governed by strong interactions. Most probably, they conserve isospin and,
therefore, $G$-parity. If so, then the $Z^\pm(4430)$ has $I^{\,G}J^P=1^+1^+$.

If widths of both initial and final mesons are taken into account, then the
$Z(4430)$ overlaps the threshold of $D_0^*(0^+)\overline{D}_1(1^+)$ (or charge
conjugate)~\cite{RPP}. One may expect that these final states are related with
the higher Fock component of the $Z(4430)$; then they will provide more intensive
modes in decays of $Z(4430)$ than $\psi^\prime\pi$ (though more difficult for
observation). It would be important to search for this decay by detailed studying
the produced system $D\bar D 3\pi$.

In terms of quark content, both $\psi^\prime\pi^+$ and $(D_0^*(0^+)
\overline{D}_1(1^+))^+$ provide evidence for the tetraquark Fock component
$\,c\bar{c}u\bar{d}\,$ in $Z(4430)^+$. If this meson contains also a canonical
quark-antiquark component, it should be $\,u\bar{d}$. Such component may develop
decays of $Z(4430)$ without any charm, with final states (m.b., through cascade
stages) of pure pion systems. Positive $G$-parity means that there should be an
even number of final pions. The lightest meson with  $I^{\,G}J^P=1^+1^+$,
$b_1(1235)$, has mass about 1230~MeV, width about 140~MeV, and decays mainly to
$\omega\pi\to4\pi$~\cite{RPP}. Having the heavier mass, $Z(4430)^+$, most probably,
produces $\geq6$ pions, at least one of which should be neutral . It seems to be
very difficult to search for such heavy resonance in such multipion system (note,
in particular, enormous combinatorial background). This was not even attempted up
to now. However, it is worth to do, since detection of the pure pion decays would
clarify the nature of $Z(4430)$, as well as other $X, Y, Z$ mesons.

\section{Conclusions}

Canonical picture of mesons as quark-antiquark systems is not self-consistent in
relaticistic theories. It can (and seems to) be, nevertheless, a good approximation
for light mesons, after appropriate adjustment of parameters for the corresponding
models. The above discussion suggests that the situation changes for heavier mesons,
where one or more higher components of a Fock column should be accounted for. It is
especially so, if the Breit-Wigner peak of the meson (or baryon as well) resonance
overlaps a threshold of pair of lighter hadrons. Then the higher Fock component(s),
with multi-quark content, may be most readily presented as a near-threshold hadron pair
in the $S$-wave state. Such a picture opens a new possible direction to investigate
the mysterious set of mesons $X,\, Y,\, Z,\,$ different from tetraquark or molecular
approaches, popular in the literature. Accurate (though difficult) experimental study
of their various decay modes can allow to separate various Fock components. To
demonstrate existence of a genuine tetraquark (without a $q\bar{q}$ component) one
should find a meson with non-canonical flavor quantum numbers, forbidden for a
quark-antiquark configuration, either appended by gluon(s) or not.

\section*{Acknowledgment}

This work was supported by the Russian Scientific Foundation, Grant No.14-22-00281.
I thank S.Afonin for a critical note.








\end{document}